# A compact, large-aperture tunable lens with adaptive spherical correction

Matthias C Wapler, Moritz Stürmer, Ulrike Wallrabe
Laboratory for Microactuators, IMTEK
University of Freiburg
Freiburg, Germany
wallrabe@imtek.uni-freiburg.de

In this paper, we present the proof of concept a very fast adaptive glass membrane lens with a large aperture/diameter ratio, spherical aberration correction and integrated actuation. The membrane is directly deformed using two piezo actuators that can tune the focal length and the conical parameter. This operating principle allows for a usable aperture of the whole membrane diameter. Together with the efficient actuation mechanism, the aperture is around 2/3 of the total system diameter – at a thickness of less than 2mm. The response time is a few milliseconds at 12mm aperture, which is fast compared to similar systems.

## I. INTRODUCTION

Adaptive lenses have been developed for a long time [1] and have in recent years been commercialized. While they can in principle replace more complex moving lens systems and can operate under conditions were moving lenses are not suitable [2], they have two major drawbacks:

On the one hand, while they can be relatively flat [3,4], they have, with the exception of electrowetting lenses [5], only a small useable aperture compared to the outer dimensions of the lens system. On the other hand, they have a fixed deviation characteristic from a spherical lens profile. Furthermore, their response times are usually a few 10s of milliseconds, depending on the size. Regarding the speed, acoustic gradient lenses are an exception, but they need to be operated in a triggered short-pulse mode [6] and are relatively bulky, with a small usable aperture. Hence, we have developed an adaptive lens with a large usable aperture/diameter ratio and a tunable aspherical behavior.

Like all fluid-membrane lenses, our proposed lens consists of a membrane that forms the lens surface and a fluid that serves as a refracting medium. Usually, the fluid is also used to exert a pressure on the membrane and hence deform it. In our case, however, the membrane is stiff and is directly deformed, such that the fluid plays no active mechanical role.

## II. OPERATING PRINCIPLE

### A. Membrane deformation modes

The active part of the lens is a thin glass membrane that is sandwiched between two piezo rings. This allows for two operating modes as shown in fig. 1:

a) A conventional bending mode, in which one piezo contracts and the other one expands (or in general both expand differently). In this case, the differential strain leads to a spherical displacement of the region of the piezo rings. This, in turn, gives a von Neumann boundary condition with a corresponding slope for the inner (passive) part of the glass membrane which bends accordingly. The piezos however do not actively produce a net radial contraction. This induces a radial stress in the glass membrane that causes a flatter deformation.

b) A buckling mode, in which both piezo rings contract, causing the stiff membrane in the middle to deflect up- or downwards. Now, the piezos essentially give a Dirichlet boundary condition to the inner part of the glass membrane. As the piezos do not bend actively, they resist the slope of the glass membrane deformation. This causes a steeper, more hyperbolic profile. In principle, this mode is bi-stable, such that direction can be chosen by the pre-deflection or an additional bending mode contribution or, more generally, by the history of the deflection.

A combination of both modes can then be used to tune the conical parameter of the surface, which in turn allows for an aspherical wavefront correction.

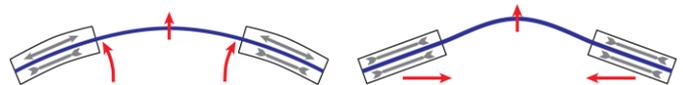

Figure 1. Actuation mechanism of the bending (left, von Neumann boundary condition) and buckling (right, Dirichlet boundary condition) mode.

One further aspect is the back pressure from the lens fluid, which will, typically, oppose the deformation and flatten the displacement. In our case, we will use a sealed but flexible fluid chamber that creates only little backpressure, such that the membrane deflection will not be affected.

## B. Theoretical estimate

We can construct a simplified geometric model to estimate the curvature parameter of the leading spherical part of the deflection. Let us consider an infinitely thin membrane and a negligible stiffness of the piezo rings. If the piezo contracts horizontally due to an applied electric field $E$ at a rate $\frac{\Delta l}{l} = d_{31}E$, $d_{31} < 0$ and the neutral planes of each piezo are separated by a distance $s$, then the inverse bending radius is approximately:

$$R^{-1} \approx s^{-1} d_{31}(E_{upper} - E_{lower}). \quad (1)$$

Typically $s$ is of the order of the piezo plus membrane thickness. Hence, typical numbers are $s \sim 100\ \mu m$, $E \sim 10^6 V/m$ and $d_{31} \sim -300 \times 10^{-12} m/V$. Interestingly, this curvature is independent of the aperture, and the focal length

$$f^{-1} = \Delta n\ R^{-1} \approx \Delta n\ s^{-1}\ d_{31}(E_{upper} - E_{lower}) \quad (2)$$

will be shorter for thinner membrane and piezo thickness. With the above values and a refractive index $n \sim 1.5$ inside and air outside the lens, this will be in the range of 100s of mm.

In the buckling mode, we consider a membrane with diameter $a$ that is compressed at its circumference with $\frac{\Delta a}{a} = d_{31}E_{mean}$. Assuming that the membrane displaces spherically and that the radial distance along the membrane is completely relaxed through the buckling, geometry leads us to:

$$R^{-1} \approx a^{-1}\sqrt{-24\ d_{31}E_{mean}}\ . \quad (3)$$

This is also interesting, as the resulting focal length

$$f^{-1} \approx \Delta n\ a^{-1}\sqrt{-24\ d_{31}E_{mean}} \quad (4)$$

predicts now a numerical aperture $a/2f$ that depends only on the piezoelectric strain and is independent of the size of the lens. Typically, the maximum value would be $NA \sim 0.1$. A negative mean electric field may either mean no deflection as the membrane is just stretched tight, or the piezo ring might buckle out of plane.

We do not investigate the conical parameter theoretically but better characterize it mechanically.

## III. REALIZATION

### A. Mechanical design

As shown in fig. 2, the lens is built on a 0.5 mm thick square-shaped glass substrate with width 19.4 mm and a supporting substrate of 0.5 mm FR2 PCB substrate. A 250 µm thick elastic polyurethane (PU) ring on top of the substrate serves as a hinge and as a spring that compensates the volume displacement of the glass membrane upon actuation. The glass membrane is 50µm thin borosilicate glass and has a slightly smaller diameter of 17mm than the piezo rings with 19mm outer and 12mm inner diameter. The latter are harvested from mass-produced acoustic transducers and have $d_{31} \approx -270 \times 10^{-12} m/V$ with 120 µm thickness. For demonstration purposes, the system is filled with 250 cSt paraffin oil with $n = 1.47$.

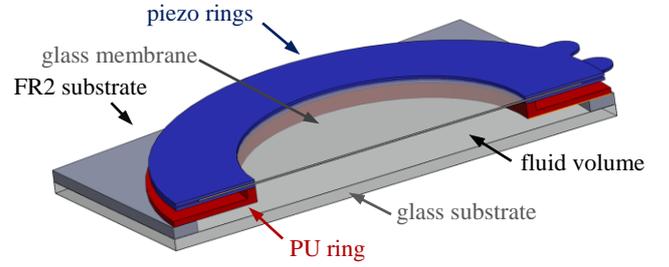

Fig. 2. Cross-section of the mechanical layout of the lens.

### B. Fabrication

The fabrication is straightforward: First, the two laser-cut piezo rings and the glass membrane are glued using Kaupo CC204 PU resin (fig. 3a). Then, the PU ring of Kaupo CF50 resin is molded directly onto the laser-cut supporting substrate using a two-piece PDMS and POM mold (b). The active part is completed by gluing the piezo-glass sandwich onto the PU ring using CF50. Finally, this body is placed upside-down and filled with oil and the glass substrate is glued on top with CF50, which still cures when immersed in oil (c).

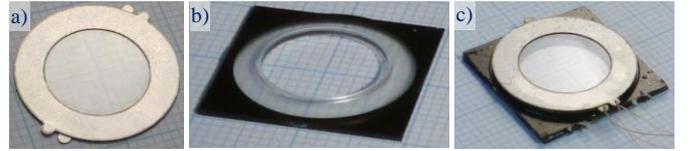

Fig. 3. Major lens components (a,b), finished lens (c).

On a production scale, this process could be performed straightforwardly using conventional pick-and-place, gluing, dispensing, and polymer molding technologies.

## IV. CHARACTERIZATION

### A. Quasistatic deformation

We performed both a mechanical and an optical characterization. To investigate the aspheric deformation of the surface, we have scanned the surface with a laser-triangulation profilometer and fitted the $r^2$ and $r^4$ components in addition to an offset and tilt. While this is, in principle, sufficient to determine the focal length and conical parameter, we also verified the focal length optically by focusing a laser beam with 4.5 mm circular aperture and measuring the beam profile along the optical axis.

In fig. 4, we show the focal length both in the bending mode up to 30 V (left) and the upwards-buckling mode up to 60 V (0.5 kV/mm, right). The 30V are the operating limit against the polarization direction (25% of the coercive field strength) and can be, in principle increased by a factor of 1.2 to 1.3, and the 60V limit were chosen to avoid any risk of electrostatic breakdown and could be increased by a factor of 2 to 4. We see that the bending mode has, up to hysteresis, an almost linear behavior. Equation (2) would give us, with $s \sim 170\ \mu m$, a focal range of $f^{-1}{}_{max} = \pm 0.2 m^{-1}$, so we have around 50% more displacement than predicted. The buckling mode has some threshold voltage and then a square-root like behavior. The former may result e.g. from an initially uneven piezo or from the force required to bend the membrane that we have neglected in the derivation. In this case, equation (4) would give us a maximum refractive power of $2.8 m^{-1}$, so also here, the experimental result is somewhat larger than predicted. The fact that the buckling mode has a much stronger refractive power comes on the one hand from the higher operating voltage and on the other hand from the different scaling of equation (2) and (4). If we had chosen thinner piezos and membranes or a larger diameter, this might be the other way round.

The optical measurement – at a static voltage on the rising branch – has a small offset to the geometric result where we used a quasi-static voltage of 3 Hz and a slightly different voltage range.

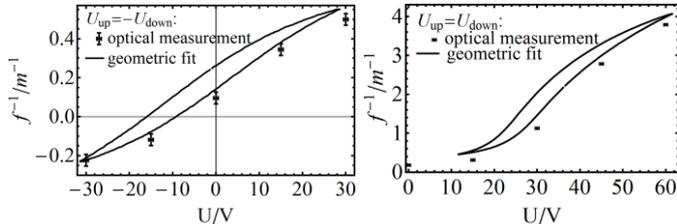
Fig. 4. Focal length in the bending mode (left) and buckling mode (right). The theoretical operating limit of the piezo material is a factor of 2 to 4 higher.

The surface unevenness was around 2.5 µm, but this was just a saddle-shaped pre-deflection that may be improved with more sophisticated fabrication. In particular, this highly depends on the unevenness of the piezo surface that can be improved by a factor of 10 (from currently 20 µm to sub-µm) for example through polishing.

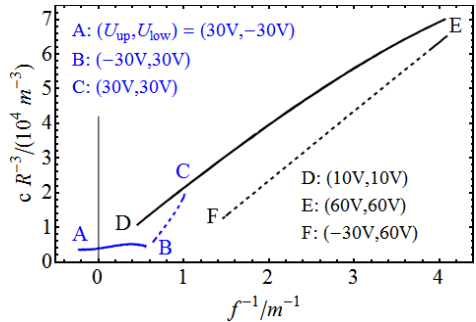
Fig. 5. Quartic parameter vs. refractive power for different actuation trajectories describing approximately the operating range for positive focal length.

In fig. 5, we show the quartic coefficient (expressed in terms of the conical parameter $c$ and curvature radius of the surface) versus the refractive power for both modes (A-B and D-E), and for a sweep from the bending to the buckling mode and vice-versa (dashed, B-C and E-F). On the one hand, the quartic parameter remained positive in all cases. On the other hand, we see that it is indeed possible to vary the conical parameter independently from the focal length. This range, bounded approximately by the trajectories may either be shifted by varying the membrane thickness and spring stiffness or it may be combined with a fixed aspherical lens to only correct the aspherical wavefront error of the total optical system that is induced by varying the focal length. The different aspherical behavior can also be seen in fig. 7, where we show 3D plots of the displacement in the bending and buckling modes.

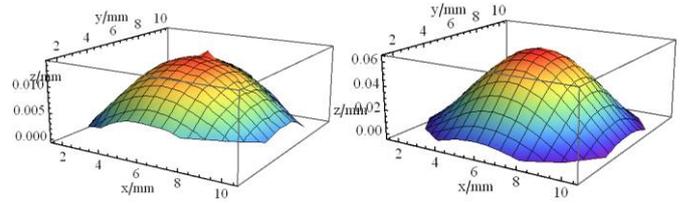
Fig. 6. Displacement in the bending (left) and buckling modes (right).

*B. Dynamic behavior*

Since the system is non-linear, we performed a perturbative frequency sweep at different pre-displacements in the buckling mode and a step response in a mixed buckling-bending mode, which are shown in fig. 7.

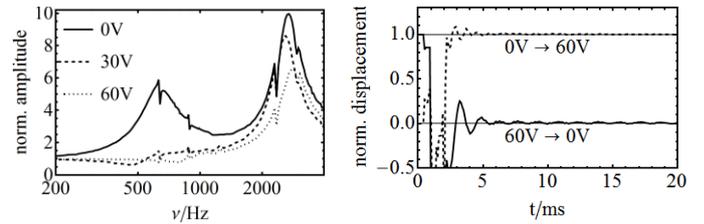
Fig. 7. Left: Resonance spectra at different pre-deflections (bias voltages). Right: Rising and falling step response in a mixed mode $(0V, 0V) \rightarrow (0V, 60V)$.

The nonlinearity manifests itself in the non-steady frequency response and in the disappearance of the first resonance mode upon pre-displacement. A Lorentzian fit to the rising side of the resonances indicates a first resonance at $\nu \approx 650\ Hz$ with a close to critical decay time of $\tau \approx 2\ ms$ and a second resonance with $2.5\ kHz$ and $1\ ms$. This is reflected in the step response, where we see a clear resonance at the falling step, but not in the rising step. This data indicates that response times around 2 to 3 $ms$ are feasible with appropriate control, which is by a factor of 10 faster than the "high speed" adaptive lens reported recently in [7]. When comparing these numbers to other adaptive lenses, one also has to keep in mind the 12 mm aperture of the system, which is larger than most lenses in the literature.

## V. Outlook and Conclusions

We have successfully demonstrated a concept for a compact aspherical adaptive lens that combines spherical wavefront correction and focusing in a single membrane. The integrated actuation mechanism allows for a compact design with large usable aperture ratio around 0.6. Furthermore, we demonstrated a fast response time of a few milliseconds at 12mm clear aperture.

Given the fact that our characterization was far below the possible maximum operating voltage of the piezo material and no optimization has yet been done, we are confident to increase the focal range towards an NA of 0.1 in future developments. Furthermore, we hope to shift the quartic parameter towards a range in the hyperbolic and elliptic region around the spherical displacement, using a suitable counter pressure.

ACKNOWLEDGMENTS

We would like to thank Nils Spengler for showing us thin glass films, which triggered this research.

This research is financed partly by the Baden-Württemberg Stiftung gGmbH project ADOPT-TOMO and funded partly by the DFG grant WA 1657/1-2 and by the BrainLinks-BrainTools Cluster of Excellence funded by the German Research Foundation (DFG, grant number EXC 1086).